\documentclass[twocolumn,showpacs,preprintnumbers,amsmath,amssymb]{revtex4}
\pdfoutput=1
\usepackage{graphicx}
\usepackage{dcolumn}
\usepackage{bm}
\usepackage{subfigure}

\begin{document}

\title{Intrinsic and extrinsic corrugation of monolayer graphene deposited on SiO$_2$}

\author{V.~Geringer$^{1,3}$, M.~Liebmann$^{1,3}$, T.~Echtermeyer$^2$, S.~Runte$^{1,3}$, M.~Schmidt$^{1,3}$, R.~R\"{u}ckamp$^{1,3}$, M.~C.~Lemme$^2$, and M.~Morgenstern$^{1,3}$}
\affiliation{ $^{1}$II. Institute of Physics, RWTH Aachen
University,\\ Otto-Blumenthal-Stra{\ss}e, 52074 Aachen\\
$^{2}$Advanced Microelectronic Center Aachen (AMICA), AMO GmbH,\\
Otto-Blumenthal-Stra{\ss}e 25, 52074 Aachen, $^{3}$ JARA:
Fundamentals of Future Information Technology}

\date{\today}

\begin{abstract}
Using scanning tunneling microscopy (STM) in ultra high vacuum and
atomic force microscopy, we investigate the corrugation of graphene
flakes deposited by exfoliation on a Si/SiO$_2$ (300 nm) surface.
While the corrugation on SiO$_2$ is long-range with a correlation
length of about 25~nm, some of the graphene monolayers exhibit an
additional corrugation with a preferential wave length of about
15~nm. A detailed analysis shows that the long range corrugation of
the substrate is also visible on graphene, but with a reduced
amplitude, leading to the conclusion that the graphene is partly
freely suspended between hills of the substrate. Thus, the intrinsic
rippling observed previously on artificially suspended graphene can
exist as well, if graphene is deposited on SiO$_2$.
\end{abstract}

\maketitle

Since it was believed, based on the Mermin-Wagner theorem, that
two-dimensional (2D) crystals are not stable at finite temperature
\cite{Mermin68}, it came as a surprise that monolayer graphene could
be stabilized on a Si/SiO$_2$ substrate
\cite{Novoselov04,Zhang05,Geim07}. Due to its peculiar properties
like, e.g., a linear dispersion leading to Klein tunneling
\cite{Katsnelson06}, high room-temperature mobility allowing quantum
Hall steps at 300 K  \cite{Novoselov07}, a low spin-orbit
interaction beneficial for spintronic devices \cite{Tombros07} or tunable spin-polarized edge states \cite{Son06},
graphene studies have become a major issue in solid-state physics.
Already the first transport results not exhibiting weak localization
lead to the speculation of a curved surface \cite{Morozov06}
acting as a phase-breaking field \cite{McCann06}. Such a curvature was
indeed observed by microscopic electron diffraction of suspended
monolayer graphene sheets \cite{Meyer07}. The lateral wavelength of
the isotropic curvature is estimated to be
$\lambda =10-25$~nm with an amplitude of $A\simeq 1$~nm. The rippling has
been reproduced theoretically by Monte-Carlo simulations with a
preferential wavelength of 8 nm barely depending on temperature
\cite{Fasolino07}. It is argued that the anharmonic coupling between
bending and stretching modes in graphene causes the rippling and is responsible for
the stability of the 2D crystal. Moreover, it has been shown experimentally
that rippling  can ultimately limit the mobility of graphene at 300 K, if defects
are avoided \cite{Morozov08}

However, the corrugation properties of graphene deposited on a substrate as
typically used in transport experiments are not clarified.
In particular, previous scanning probe studies
\cite{Stolyarova07,Ishigami07,Tikhonenko08,Zhang08} revealed only
corrugations, which are attributed to the roughness of the underlying
substrate, except for a torn flake manipulated by AFM \cite{Tikhonenko08}. Thus, it is unclear, if the intrinsic tendency for rippling
persists on the substrate, thereby ultimately limiting mobility and influencing
weak localization, which has meanwhile been found, at least, for some samples
\cite{Tikhonenko08}. Here, we demonstrate that a regular short wave-length
corrugation with $\lambda\simeq 15$~nm and $A\simeq 1$~nm, which is
not induced by the substrate, can even prevail the substrate corrugation.
Since $\lambda$ and $A$ are close to the values
found on suspended graphene
\cite{Meyer07}, we attribute this additional corrugation to the intrinsic rippling
of graphene \cite{Fasolino07}.
Moreover, we find that the long-range corrugation implied by the substrate
has lower amplitude on graphene than on SiO$_2$ suggesting that our high-mobility
graphene is partly suspended between hills of the substrate, which
might favor the development of the intrinsic short-scale rippling.

\begin{figure}
\includegraphics[width=\linewidth]{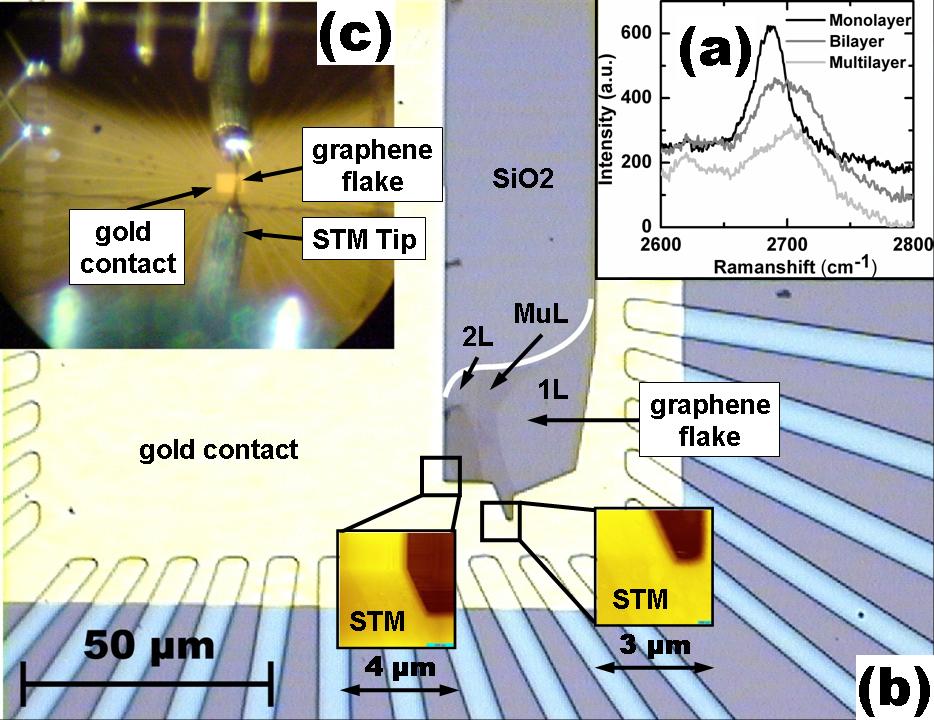}
\caption{\label{fig1} (Color online) (a) Raman spectra using laser light of wavelength 532.1~nm.
The spectra are recorded on different areas of the graphene flake and the attribution to monolayer, bilayer and multilayer is marked.
(b) Optical image of the graphene sample with gold contacts. The different areas of monolayer (1L), bilayer (2L) and multilayer (MuL) as identified by Raman spectroscopy are indicated. Insets show two STM images acquired at the edge of the gold contact, which are used for orientation. (c) Microscopic image of the STM tip in tunneling contact with the graphene in UHV recorded with a long-range optical microscope.}
\end{figure}

The graphene sample is fabricated by the mechanical exfoliation
technique as described in \cite{Novoselov04,Lemme07}. Using an
optical microscope, a graphene flake containing a monolayer region
is identified. Raman spectroscopy is used to confirm the number of
layers with the help of the 2D line \cite{Ferrari06}.
Fig.~\ref{fig1}(a) shows three spectra measured in different areas
of the flake visible in Fig.~\ref{fig1}(b). In addition to a large
monolayer area, we find smaller bi- and multilayer areas as marked
by 1L, 2L and MuL, respectively. Next, gold contacts with a 10~nm Cr
seed layer were deposited and structured with a lift-off process.
Except for one side, the graphene sample is completely surrounded by
the gold electrode as shown in Fig.~\ref{fig1}(b). The mobility of
graphene samples prepared identically but with several contacts
is $\mu=1-1.5$ m$^2$/(Vs) at 300 K \cite{Lemme08}, i.e. comparable with high-mobility
values obtained by other groups \cite{Morozov08,Tan07}. In order to
remove residual resist and adsorbates, the sample was rinsed in
isopropanol and acetone, baked out to 150$^\circ$ C in air for four
hours and, additionally, baked out to the same temperature in
ultrahigh-vacuum (UHV) for three hours.
The gold film served as the electrical contact for scanning tunneling microscopy (STM) and was used to prepare the STM tip (etched W) by voltage pulses.
In order to find the graphene within the UHV-STM, we used an optical
microscope with a focal length of 30~cm and 5~$\mu$m
lateral resolution. Fig.~\ref{fig1}(c) shows a microscopic image of
the tip approaching the graphene flake. The home-built STM
features an $xy$ stage for lateral positioning and operates in UHV
($2\cdot 10^{-7}$ Pa) as described elsewhere
\cite{Wiebe04,Geringer08}. The $xy$ stage has been checked to move
accurately within 10 \%. Topographic images are recorded applying
a bias $V$ to the tip. Spectroscopic $dI/dV$ curves are
measured by lock-in technique using a modulation voltage $V_{\rm
mod}$ after stabilizing the tip at voltage $V_{\rm stab}$ and
current $I_{\rm stab}$. STM images
obtained at the intentionally irregular edge of the gold contact are
displayed in the insets of Fig.~\ref{fig1}(b). They are used for
orientation by STM and aid in finding monolayer, bilayer and
multilayer areas on the flake.

Atomic force microscopy images are taken under ambient conditions in
tapping mode using either a commercial cantilever with a silicon tip or an
ultrasharp tungsten tip with a tip radius of 1~nm, which is attached at the
bottom of a Si cantilever \cite{cantilever}. Imaging has been carried out in the attractive regime \cite{holscher07} with an oscillation frequency slightly above resonance.

\begin{figure*}
\includegraphics[width=\linewidth]{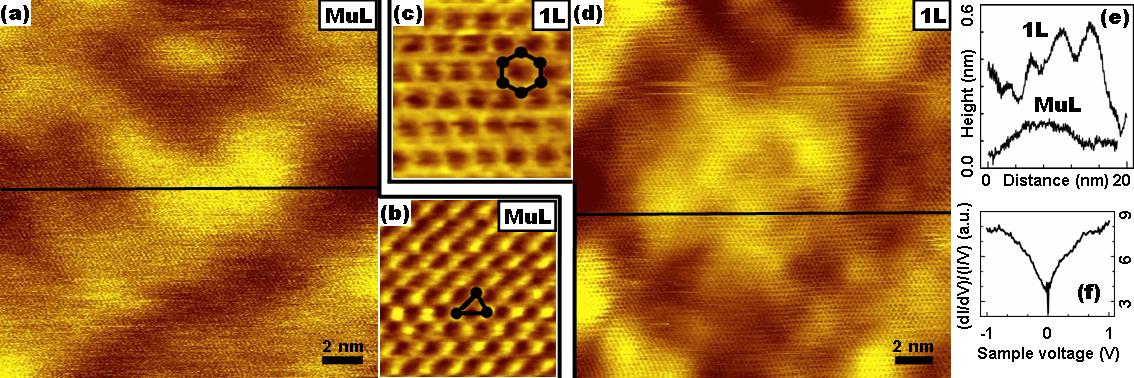}
\caption{\label{fig2} (Color online) (a) Constant current STM image
of multilayer graphene (0.4\,V, 258\,pA) (raw data). (b) Higher
resolution STM image of the same area as in (a) (0.4\,V, 198\,pA).
(c) High resolution STM image of graphene monolayer area (1\,V,
394\,pA). (d) Lower resolution STM image of monolayer graphene
(0.5\,V, 292\,pA) (raw data). (e) Line section along the lines
depicted in (a)(MuL) and (d)(1L). (f) Normalized dI/dV spectrum of
monolayer graphene: $V_{\rm stab}$=0.7\,V, $I_{\rm stab}$=300\,pA,
$V_{\rm mod}$=20\,mV.}
\end{figure*}

Fig.~\ref{fig2}(b) and (c) show two atomically resolved STM images
recorded on the monolayer (1L) and the multilayer (MuL) of graphene,
respectively. One observes a hexagonal pattern on the monolayer and
a triangular pattern on the multilayer in accordance with previous
results \cite{Stolyarova07, Zhang08}. The atomic resolution is still
perceptible at lower resolution in Fig.~\ref{fig2}(a) and (d), but
an additional irregular corrugation appears. Line sections shown in
Fig.~\ref {fig2}(e) reveal that the corrugation height on this
monolayer region is about 0.6~nm, a factor of three larger than on
the multilayer region. Notice that corrugation heights pretended by
charged defects are typically smaller by more than an order of magnitude \cite{Wittneven98}.
A representative $dI/dV/(I/V)$ curve of the
monolayer, which is known to represent the local density of states
(LDOS) \cite{Stroscio86}, is shown in Fig.~\ref{fig2}(f). As
expected for graphene, it shows a V-like shape with the Dirac point
at about $V=- 20$ mV.
It does not change significantly across the area depicted in
Fig.~\ref{fig2}(d). Note that graphene on SiC(0001)
exhibits a more complicated $dI/dV$ spectrum
\cite{brar07,rutter07,lauffer08}. Note also, that we
did not observe the phonon gap found at 4 K recently \cite{Zhang08},
although we used several different micro- and macrotips. This is tentatively attributed
to the different temperature of the two experiments.
The good representation of the graphene LDOS by $dI/dV/(I/V)$ and the continuous atomic resolution across Fig.~\ref{fig2}(d) demonstrates the absence of resist on this part of the surface. Indeed, we found several large areas without resist, but still also areas where a remaining resist is apparent within the STM images.
The fact that the Dirac point is observed
close to 0 V shows a negligible influence of charging adsorbates,
which are probably removed in UHV.

\begin{figure*}
\includegraphics[width=\linewidth]{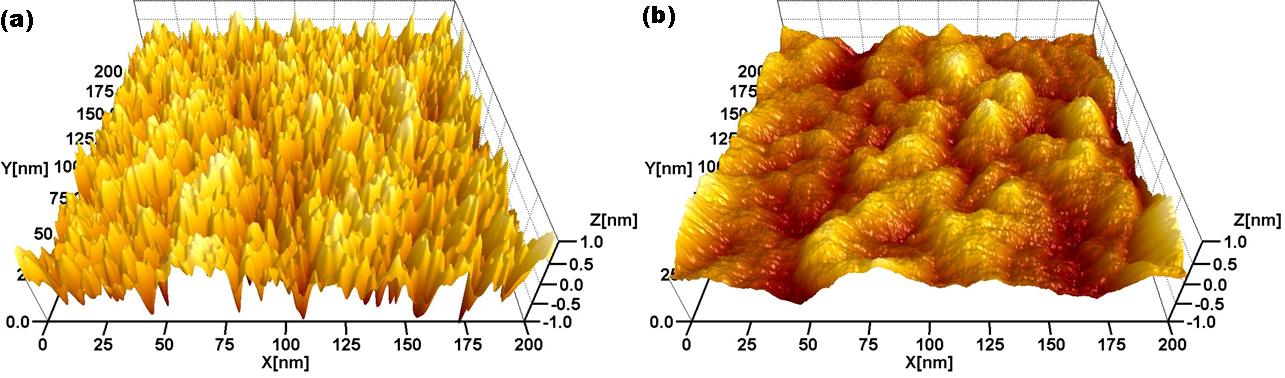}
\caption{\label{fig3} (Color online) (a), (b) 3D and 2D constant current STM image of
monolayer graphene (1\,V, 207\,pA). (c) 3D tapping-mode AFM image of the ${\rm SiO_2}$ substrate (resonance frequency 326.4~kHz, force constant 47~N/m, excitation frequency 326.5~kHz, oscillation amplitude 18~nm, constant amplitude feedback, setpoint 90\%).}
\end{figure*}

Fig.~\ref{fig3}(a) shows a large area STM image of the graphene
monolayer. One observes a rather regular corrugation with amplitudes
of more than 1~nm and a preferential distance between hills of about
15~nm. The histogram \cite{EPAPS} is Gaussian with a full width at
half maximum (FWHM) of 0.78~nm. The rms roughness is determined to
be 0.36~nm. The latter two values fluctuate across the monolayer
area by about 25 \% (average rms roughness 0.32~nm), but the
corrugation length scale remains constant. For comparison,
Fig.~\ref{fig3}(c) shows an AFM image of the bare SiO$_2$ surface
recorded in tapping-mode. It is obvious that the corrugation on the
substrate is lower and exhibits a larger length scale than the
corrugation on graphene. The FWHM of the histogram of
Fig.~\ref{fig3}(c) is 0.48~nm and the rms roughness is 0.22~nm
(average of all data $(0.25 \pm 0.05)$~nm). We used four different
tips, including two ultrasharp tips with a tip radius of 1~nm
\cite{cantilever}, but always observed the same corrugation height
and length scale, which, in addition, is in good agreement with
previous results \cite{Ishigami07}.

\begin{figure}
\includegraphics[width=\linewidth]{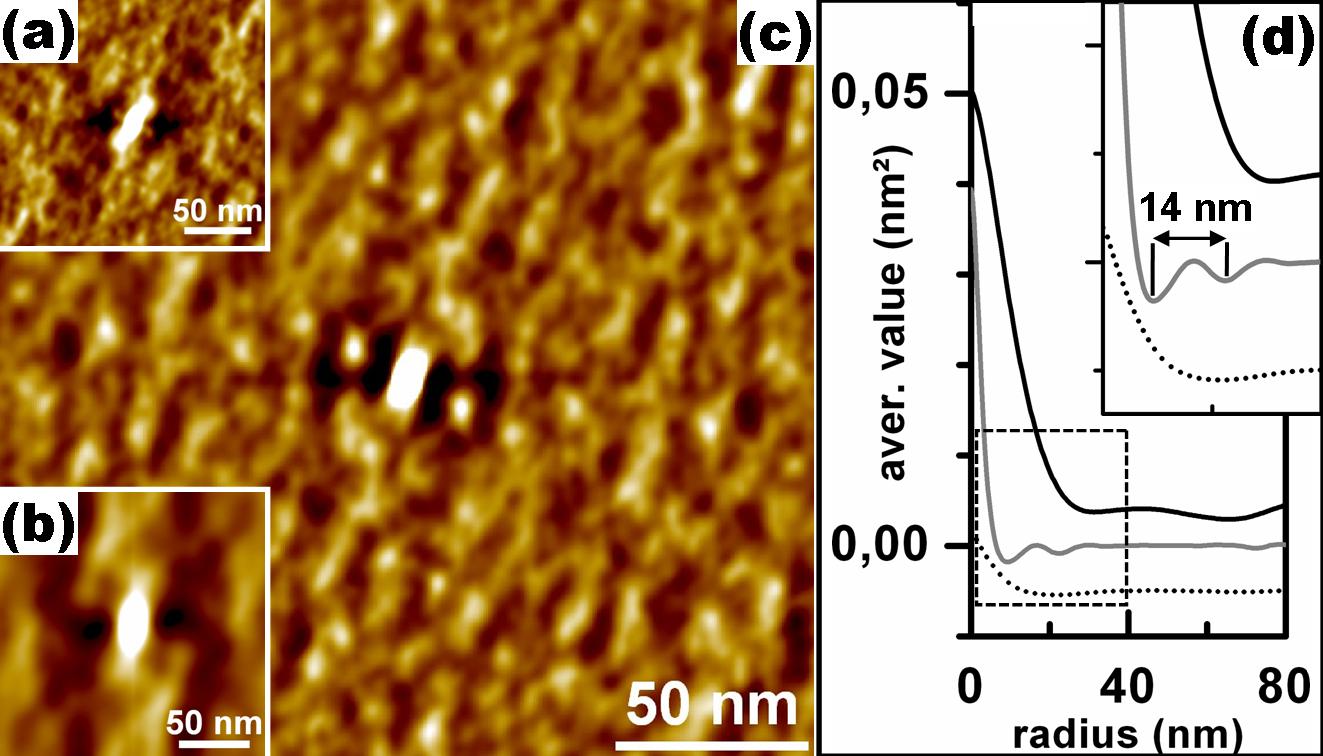}
\caption{\label{fig4} (Color online) (a) 2D autocorrelation function of the
drift corrected STM image of monolayer graphene as shown in
Fig.~\ref{fig3}(a). (b) 2D autocorrelation function of the AFM image
of the SiO$_2$ surface as shown in Fig.~\ref{fig3}(c), same color scale as in (a). (c) FFT high
pass filtered image (1.\ order Butterworth, 20\,nm) of the 2D autocorrelation
function shown in (a). (d) Radial line sections of the unfiltered AFM autocorrelation image (b) (black), of the high pass filtered STM autocorrelation image (c) (gray), and of the high pass filtered AFM autocorrelation image (dotted). The inset shows a larger view of the area marked by the dashed line. The black and gray graphs are offset vertically for clarity.}
\end{figure}

Careful inspection of the graphene monolayer in Fig.~\ref{fig3}(a) and (b)
shows a modulation on top of the short-range corrugation exhibiting
a similar length scale as the corrugation on the SiO$_2$ substrate.
In order to disentangle the two contributions, we used
two-dimensional autocorrelation functions of the images (a) and (c)
\cite{horcas07}, which are shown in Fig.~\ref{fig4}(a) and (b).
Note that both figures use the same color scale, i.\ e.\ the
intensities are directly comparable. A long-range structure
represented by a central spot surrounded by four bright areas is
visible in both images although with weaker intensity for graphene
(a). Fig.~\ref{fig4}(c) shows the autocorrelation function of the
graphene after removing the long-range part by high-pass filtering
using a smooth (first order Butterworth) cutoff at wavelength
$\lambda=20$~nm. A preferential short-range distance is clearly
visible by the strong spots around the center, but there is
additional multiple correlation albeit not with high symmetry.
Notice that creep and drift effects are carefully removed from the
STM-image using the atomic resolution. We checked the preferential
directions visible in both correlation patterns. They are rotating arbitrarily across the SiO$_2$, but we find indication for a preferential orientation on graphene \cite{EPAPS}.
Fig.~\ref{fig4}(d) shows
radial line sections averaged over all angles of the correlation
functions. The SiO$_2$ (black curve) exhibits a correlation length
of about 25~nm comparable to \cite{Ishigami07} and, in addition, a
very weak maximum at about 50~nm indicating a slight preferential
distance between hills. The gray curve shows the radial line section
of the high-pass filtered image of graphene visible in
Fig.~\ref{fig4}(c). It exhibits a damped oscillation with a wave
length of about 14~nm looking very similar to correlation functions
of liquids. For comparison, the radial line section of the
autocorrelation function of the AFM image after applying the same
high-pass filter is shown as a dotted line and does not exhibit any
structure. We observe the damped oscillation also after high-pass
filtering the original STM image in Fig.~\ref{fig3}(a) and
determining the autocorrelation function afterwards. In addition, we
checked that the obtained wave length is quite robust with respect
to slight parameter changes of the filtering process. Depending on
the filtering procedure and the selection of the minima or maxima
used for determining the wave length, $\lambda$ is varying by at
maximum $\pm 2$ \,nm with a mean value of $\lambda = 15$~nm.

From these results, we conclude that monolayer graphene can exhibit
a rather regular, liquid-like short-scale corrugation not induced by the
substrate. It is similar in height and wavelength to the one
observed on suspended graphene \cite{Meyer07} and calculated for
freely suspended graphene by atomic Monte-Carlo methods
\cite{Fasolino07}. This observation suggests that the graphene might be
partly suspended even on the SiO$_2$.
In order to verify this, we compared the Fourier components of the
images of graphene and SiO$_2$ within the large wavelengths region
$\lambda =40-60$~nm \cite{EPAPS}. We indeed find that graphene
exhibits an  up to 40~\% lower amplitude at all wavelengths within
this wavelength range. This implies that the graphene does not
follow the substrate corrugations exactly, but instead includes
areas not in contact with the substrate. We believe that this partly
free-standing configuration is the prerequisite for the intrinsic
rippling of graphene. Indeed, we have also found other graphene
flakes prepared nominally identically by the same person which did
not show the intrinsic rippling. This highlights that tiny details
of the preparation process can have large impacts on the physical
properties, maybe explaining the contradicting weak localization
properties in different studies \cite{Tikhonenko08,Morozov06}. A
comparison of Raman data of several samples indicates that samples
with intrinsic rippling are more frequent than samples without
\cite{EPAPS}. Future systematic studies using our method to
determine the intrinsic rippling are required.

In summary, we have investigated high-mobility graphene flakes on SiO$_2$ by
UHV-STM and scanning tunneling spectroscopy. A mesoscopic corrugation decreasing in
amplitude with increasing thickness
not induced by the substrate is identified. It is rather regular exhibiting liquid-like
correlation properties with a preferential wavelength
of about 15~nm and a rms roughness of 0.32~nm. Since the long-range
corrugation of the substrate is additionally visible on graphene,
but with smaller amplitude than on the substrate, we conclude that
the rippled graphene is partly freely suspended.

We gratefully acknowledge useful
discussions with G. G\"untherodt, U. D. Schwarz and H. Kurz and financial support by FOR
912, project TP 6 of the Deutsche Forschungsgemeinschaft and by the
German Federal Ministry of Education and Research (BMBF) under
contract number NKNF 03X5508 ("ALEGRA").

\section*{Supplementary information}

\subsection*{Histogram of Corrugation}
\begin{figure}
\includegraphics[width=\linewidth]{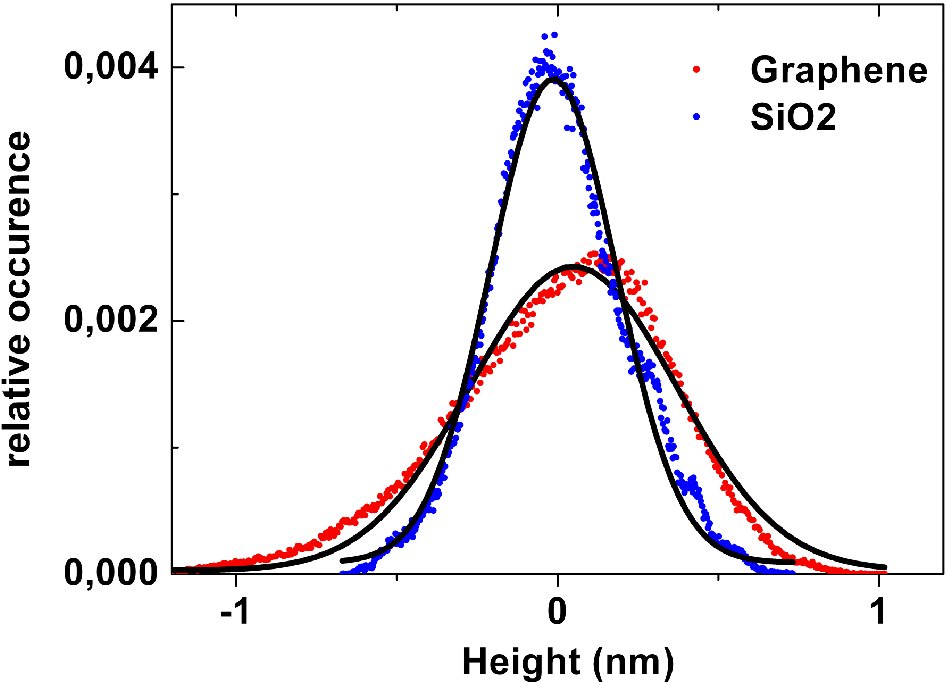}
\caption{\label{figs1} (Color online) Height histograms of SiO$_2$ (blue) and graphene (red) corresponding to the images in Fig.\ 3 of the main text. On average, the graphene histogram exhibits a larger FWHM value compared to SiO$_2$.}
\end{figure}

Figure \ref{figs1} shows representative histograms of the height values on the SiO$_2$ substrate
and the graphene sample as obtained
by atomic force microscopy (AFM) and scanning tunneling microscopy (STM), respectively. The histograms are based on
images of 200 $\times$ 200 nm$^2$ including 512 $\times$ 512 measurement points each. These images are shown in
Fig. 3 of the main text. The histogram curves are normalized in
order to enclose the same integral area. Gaussian fits are added as guides to the eye. Obviously, the
height histogram of Graphene is broader exhibiting a full width at half maximum (FWHM) of 0.78 nm, while
SiO$_2$ exhibits a FWHM of only 0.48 nm. While the latter value is quite robust across the surface, the former
value is changing by about 25 \% across the surface, but is on average larger than the FWHM on SiO$_2$.
This demonstrates that the rippling of graphene is not limited to substrate corrugations.

\subsection*{Tip Dependence of Substrate Corrugation}
\begin{figure*}
\includegraphics[width=\linewidth]{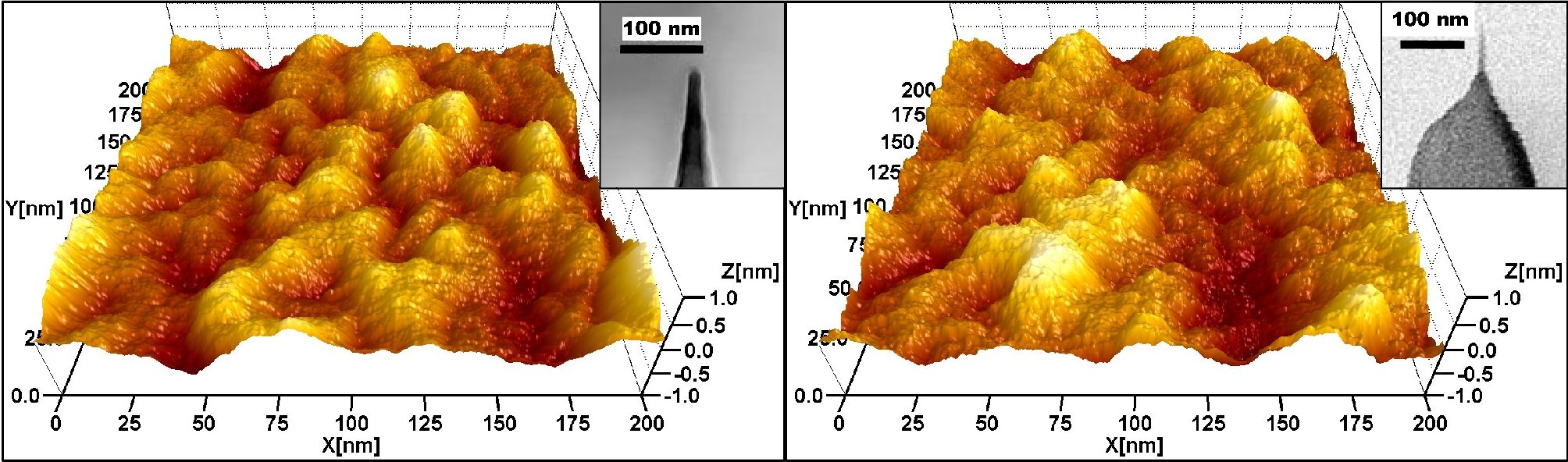}
\caption{\label{figs2} (Color online) AFM images of SiO$_2$ with tips of different radii: Si tip NSC15 with 10 nm (a) and W/Si tip DP15/HiRes-W/AlBS with 1 nm (b), respectively. Both images show the same color scale. Insets: SEM images of similar tips, with kind permission of MikroMasch (www.spmtips.com).}
\end{figure*}

Figure \ref{figs2} shows two AFM images of the SiO$_2$ substrate obtained in tapping mode by two different tips.
Representative scanning electron microscopy (SEM) images of the tips provided by the
supplier are added as insets. They are cross-checked by SEM images after the measurements. These images are in agreement
with the supplier images,
but due to a moderate background pressure, the ultrasharp W-tip fastly burnt away during SEM imaging.
The left AFM image is measured by a cantilever with a Si tip having an apex radius of about 10 nm, while the right AFM image is measured
with a W tip of nominal tip apex radius of 1 nm mounted at the bottom of the Si tip. Both images exhibit a very similar lateral scale of the corrugation.
The rms values are 220 pm and 270 pm, respectively, showing that the tip properties do not significantly change the
observed corrugation of the substrate. We cross-checked this result using more than ten images obtained with both tips, which lead to fluctuations in rms values of less than $\pm$ 10\%.

\subsection*{Comparison of Correlation Functions}
\begin{figure*}
\includegraphics[width=\linewidth]{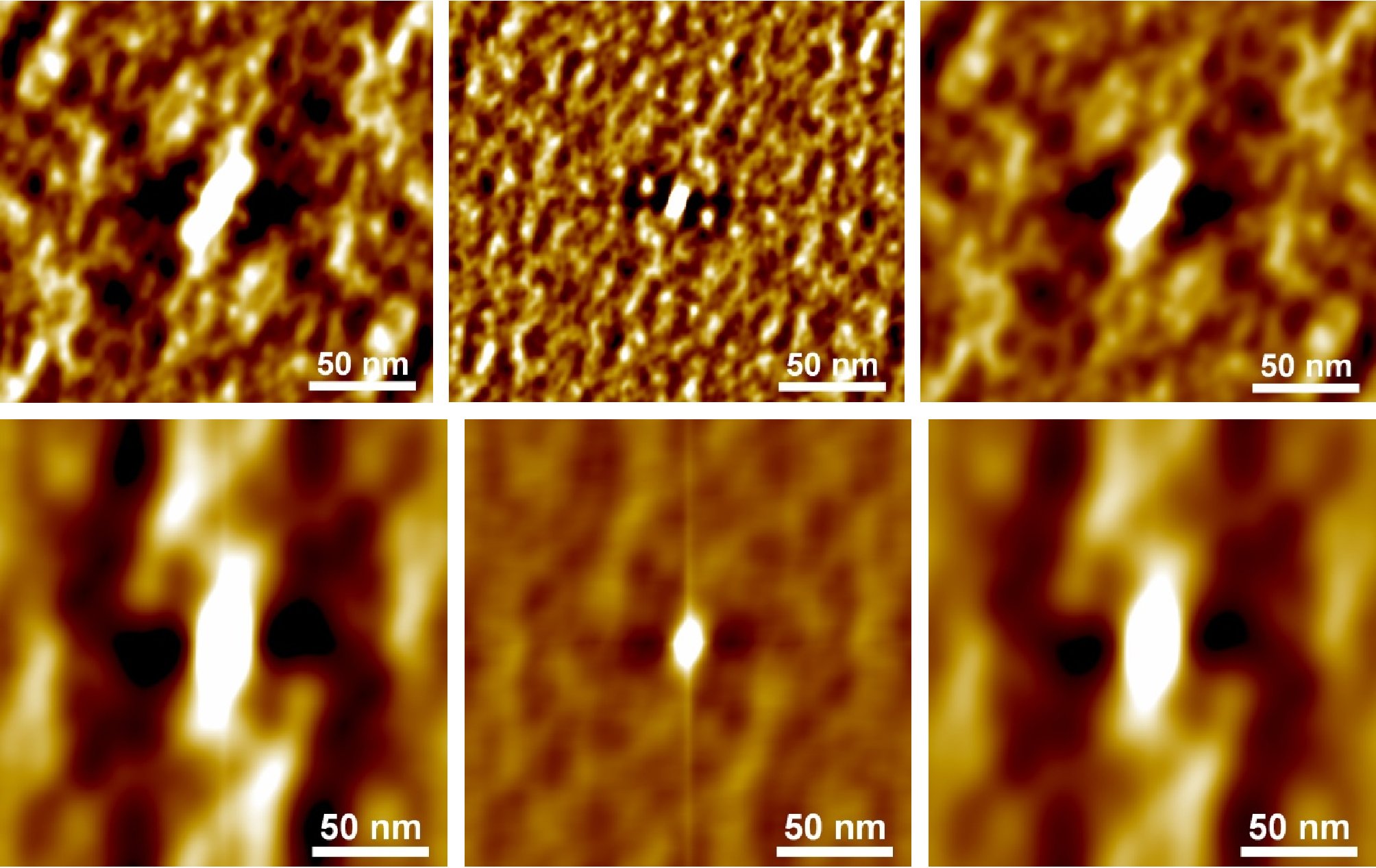}
\caption{\label{figs3} (Color online) 2D autocorrelation images of graphene (top row) and SiO$_2$ (bottom row). Left: autocorrelation calculated from original images. Middle and right: high-frequency and low-frequency part, obtained by high pass or low pass filtering, respectively, using a 20 nm first-order Butterworth filter. Note that the color scale is identical within each column. Drift effects have been removed from the graphene images by using the atomic resolution.}
\end{figure*}

Figure \ref{figs3} shows a comparison of the correlation functions of the topographic images of graphene (top row) and SiO$_2$ (bottom row).
The left column shows
the correlation functions directly obtained from the topography. The middle and right column show the high-frequency
and low-frequency part of the correlation function separated by a smooth first-order butterworth filter with cutoff
wavelength $\lambda=20$ nm. While the large wavelengths visible in the right columns are quite similar in both images, the rather regular short wavelength pattern (middle column) is only observed on graphene. Note that correlation functions within the same columns are scaled to the same contrast, i.\ e., their brightness is directly comparable.

\subsection*{Reproducibility of Correlation Functions}
\begin{figure*}
\includegraphics[width=\linewidth]{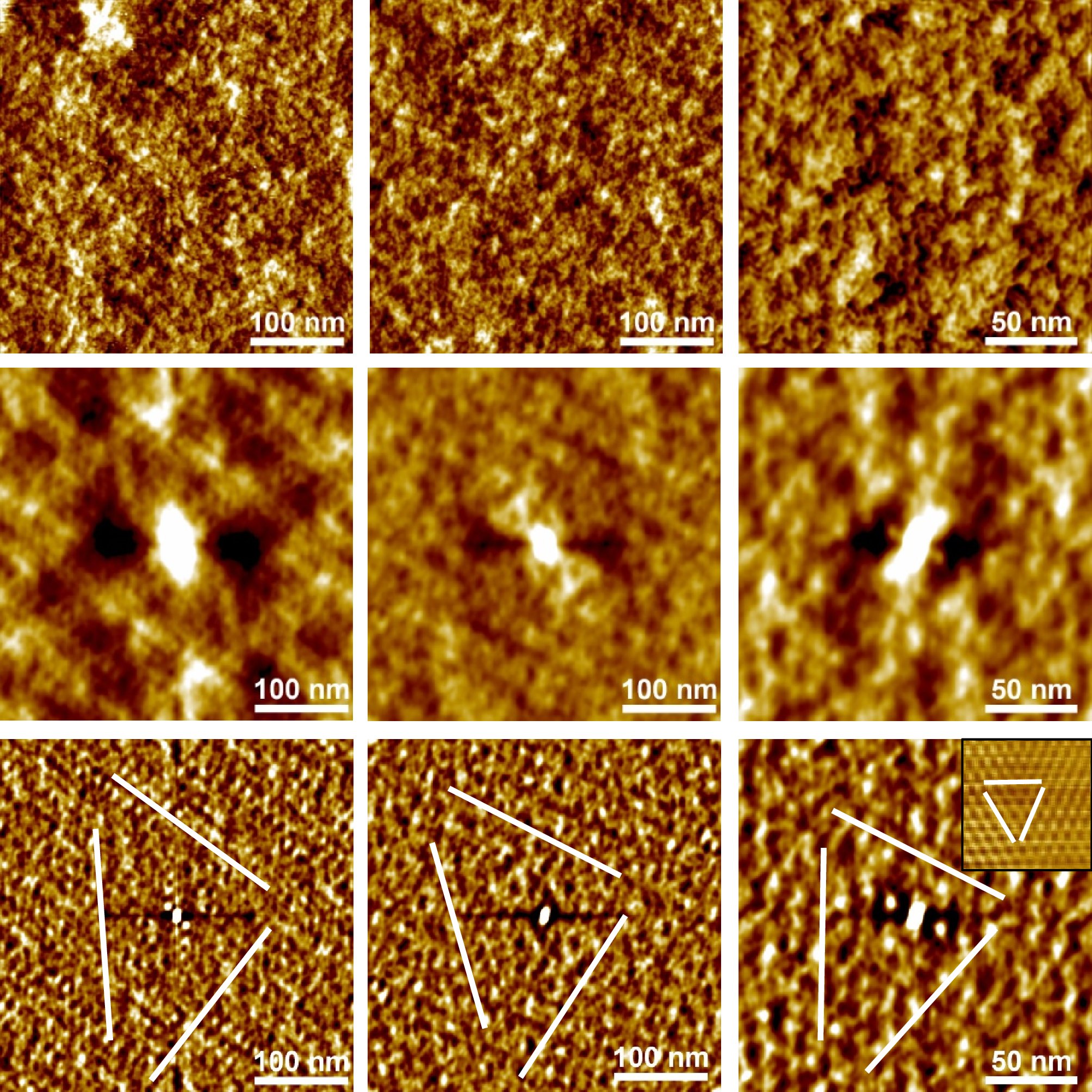}
\caption{\label{figs4} (Color online) Different graphene
areas (top row), separated by several $\mu$m, with autocorrelation images (middle) and high-pass filtered autocorrelation images (bottom row). The preferential orientation of long-range corrugation varies arbitrarily, while the short-range corrugation shows similar orientation. Inset: Autocorrelation of the atomic resolution image.}
\end{figure*}

Figure \ref{figs4} shows three images (top row) obtained in different areas of the monolayer graphene, separated by several $\mu$m.
They exhibit different rms values of 270 pm, 220 pm and 360 pm (left to right), respectively. The middle row shows the
corresponding correlation functions. The bottom row shows the same correlation functions after high-pass
filtering using the identical procedure as in Figure 3. Notice that the scales of the images are different.
While the long-range corrugation (middle row) is oriented in different directions, the short range corrugations
(bottom row) exhibit the same
wave length of about 15 nm in all three areas and show a similar orientation as marked by the white lines. The latter indicates that the the short-scale corrugation is related to the atomic directions in graphene. But since the directions of the rippling are fluctuating slightly across the sample similar to the orientations of molecules in liquid crystals, more data are required in order to substantiate the relation between atomic directions and preferred directions of rippling. 

\begin{figure*}
\includegraphics[width=\linewidth]{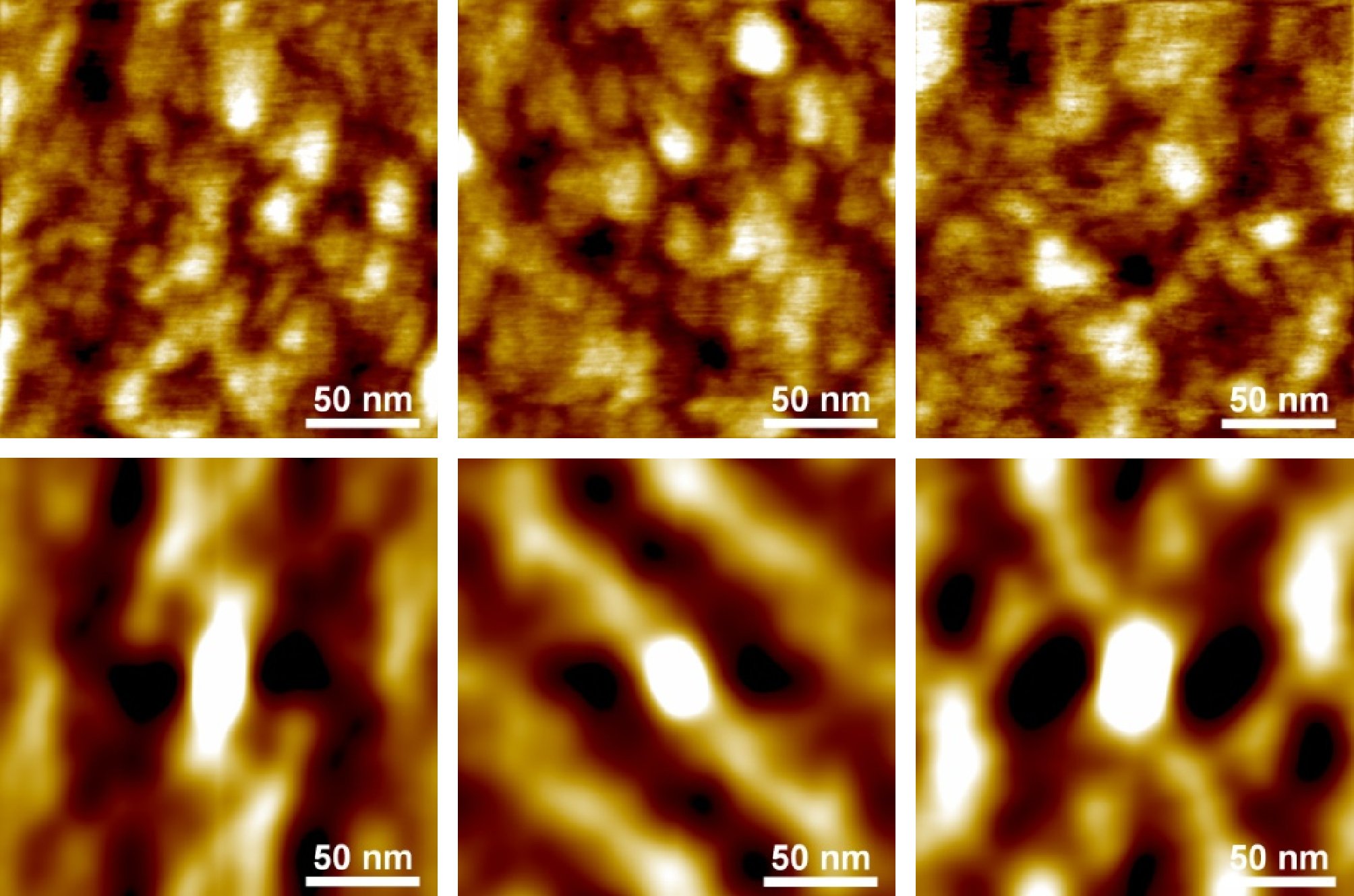}
\caption{\label{figs5} (Color online) AFM images of different SiO$_2$ areas (top) with corresponding 2D autocorrelation (bottom). The preferential orientation varies arbitrarily.}
\end{figure*}

Figure \ref{figs5} shows topographic images of different areas of the SiO$_2$ substrate (top row) and the corresponding correlation functions
(bottom row). They exhibit similar length scales of corrugation but clearly different orientations, which might be due to the polishing procedure of the sample.

\subsection*{Wave Length Dependence of Corrugations}
\begin{figure}
\includegraphics[width=\linewidth]{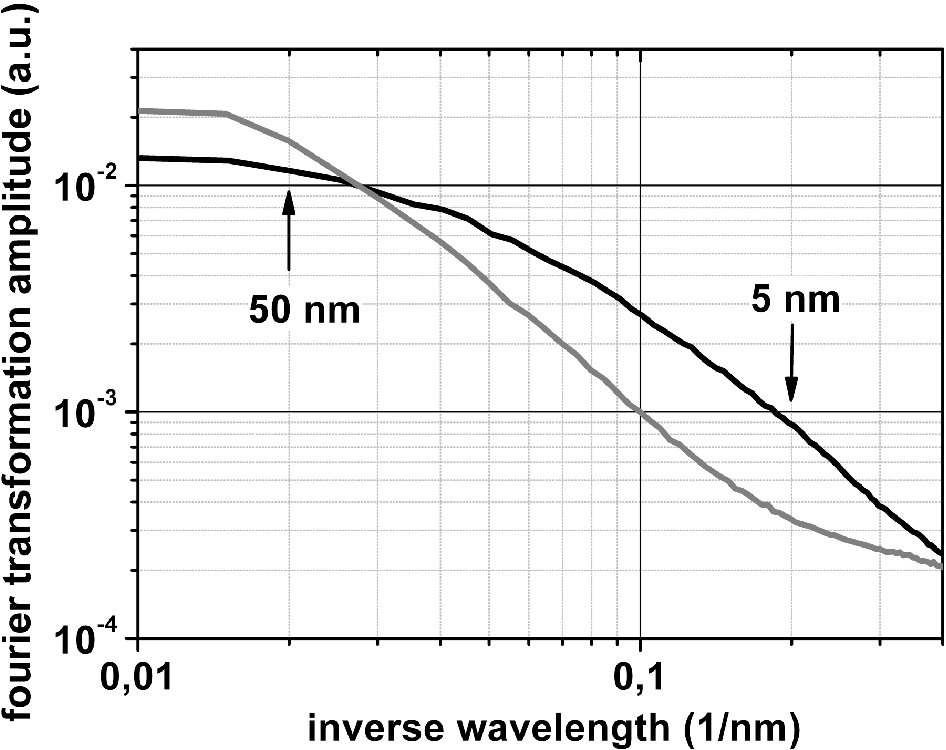}
\caption{\label{figs6} Radial averages of the Fourier transformations of graphene (black) and SiO$_2$ (grey) images, averaged over 10 images. The crossing point of individual curves varies between 20 and 40 nm.}
\end{figure}

Figure \ref{figs6} shows the radially averaged Fourrier transforms (FT) of the topographic images of graphene (black curve) and SiO$_2$ (grey curve). The FT curves are based on images
of the same lateral size and resolution. The curves are averaged using 10 images of 200 $\times$ 200 nm$^2$ from different areas of the surfaces. Obviously,
the corrugation of SiO$_2$ is larger at large wave length (small wave number) down to 30 nm by up to 66 \%, while the corrugation on the
graphene is larger at smaller wavelengths (larger wave numbers) by up to a factor of 2.6. The individual FTs of different SiO$_2$ areas are nearly indistinguishable, i.e they are not deviating from the width of the grey curve in Fig. 5. However, the FTs of different graphene areas are shifted vertically due to the fluctuating rms value of corrugation, which leads to crossing points of
the two curves varying between wavelengths of 20 nm and 40 nm. Importantly, all of the graphene curves show one
crossing point with the averaged SiO$_2$ curve.

\subsection*{Raman investigations on monolayer graphene}

\begin{figure*}
\includegraphics[width=\linewidth]{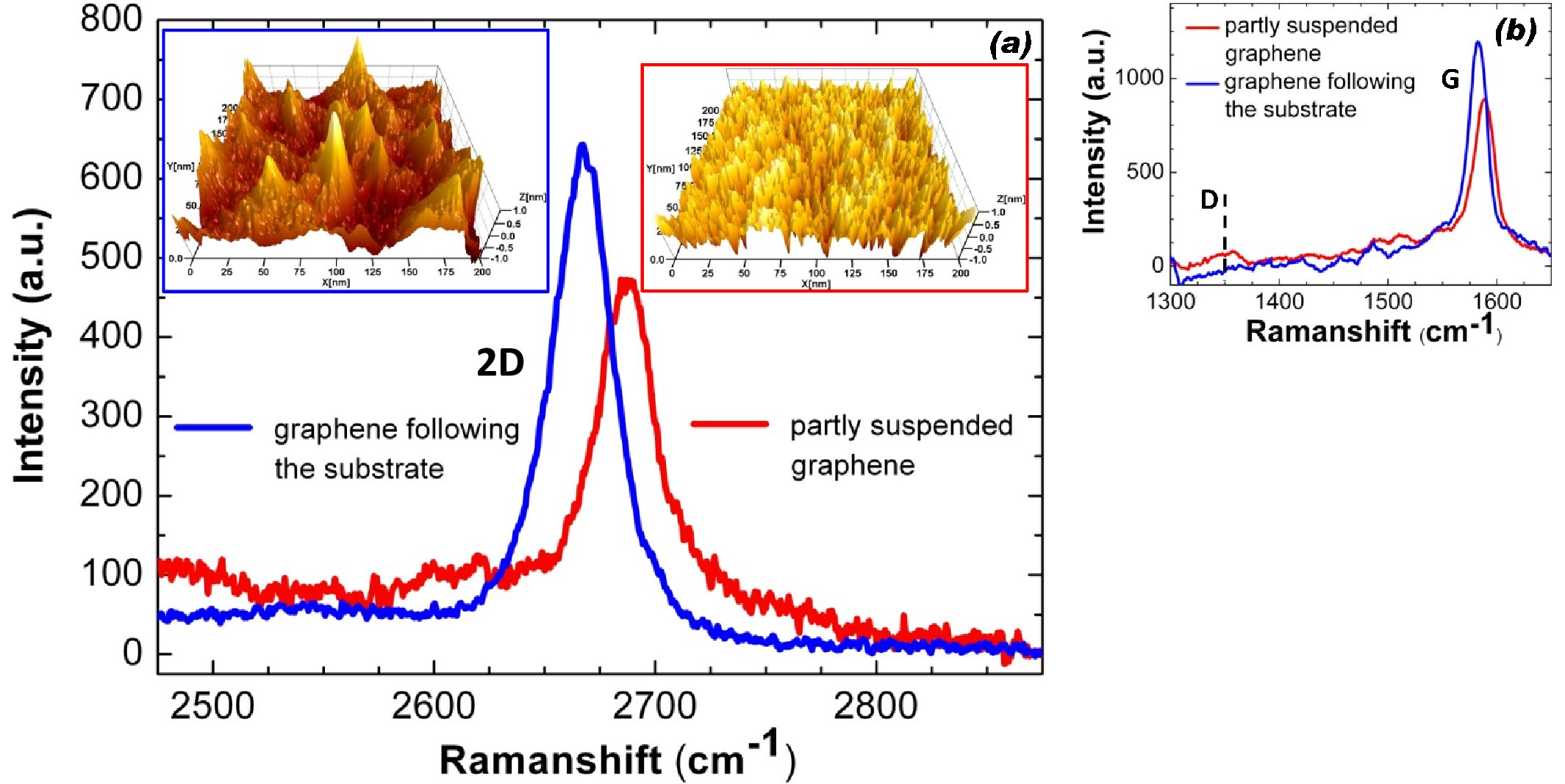}
\caption{\label{figs7} (Color online) (a)  Comparison of Raman
spectra at 532.1 nm laser wave length for partly suspended and
following the substrate graphene  monolayer flake.  The 2D peak
position is 2686.7 cm$^{-1}$ for partly suspended graphene and
2670.5 cm$^{-1}$ for the flake following the substrate. The
experimental conditions for both samples are identical. Left inset:
STM measurement of the graphene flake following the substrate (0.4
V, 0.2 nA). Right inset: STM measurement of the partly suspended
graphene flake (1 V, 0.2 nA). (b)  Raman spectra of the D and G band
 for the same samples. The D line intensity (1350 cm$^{-1}$, marked by a dashed line) is very weak for both
samples. The G line from the flake following the substrate is
shifted to lower values by 5.5 cm$^{-1}$ with respect to the G peak
of the partly suspended graphene.}
\end{figure*}

In order to observe the influence of the graphene film morphology
on Raman scattering, we compared Raman spectra (laser wave length
532.1 nm), taken in identical experimental conditions, of a
monolayer flake which is partly suspended and one that follows the
substrate (Fig.\ \ref{figs7}). The surface structure of these samples was
determined using STM measurements (Insets Fig.\ \ref{figs7}a). The Raman
spectra were recorded before and after the STM measurements giving
the same result. A difference between both spectra regarding the
positions of the 2D and G peaks is clearly visible. The wave number
of the 2D peak is 2686.7 cm$^{-1}$ for the partly suspended graphene
flake and 2670.5 cm$^{-1}$ for the flake following the substrate.
Hence, it results that the change in morphology causes a shift of
the 2D peak by 16.2 cm$^{-1}$ (Fig.\ \ref{figs7}a). A similar effect was
observed at the G peak which is shifted by 5.5 cm$^{-1}$ to lower
values  for the flake following the substrate (Fig.\ \ref{figs7}b). The D peak
(about 1350 cm$^{-1}$) is barely visible for both samples indicating
a good quality of our graphene. Raman spectra of several of our
graphene samples (about 15) prepared in nominally identical
conditions show more frequently a non shifted 2D line with an
average value of 2692.4$\,\pm\,$5.3 cm$^{-1}$. Hence, we conclude
that our samples mostly exhibit intrinsic rippling. If we compare
with wave numbers of the 2D line within the literature using only
the data taken with a similar laser wave length (514 - 532 nm)
\cite{Ferrari06,Gupta06,Graf07}, we find 2680 - 2700
cm$^{-1}$. Hence, we believe that the majority of samples
investigated so far exhibit intrinsic rippling as displayed in the
right inset of Figure 7a.

\end{document}